\documentclass[conference]{IEEEtran}
\IEEEoverridecommandlockouts
\usepackage{cite}
\usepackage{amsmath,amssymb,amsfonts}
\usepackage{algorithmic}
\usepackage{graphicx}
\usepackage{textcomp}
\usepackage{xcolor}
\usepackage{multirow}
\usepackage{subcaption}
\usepackage{url}
\usepackage{epstopdf}
\def\BibTeX{{\rm B\kern-.05em{\sc i\kern-.025em b}\kern-.08em
    T\kern-.1667em\lower.7ex\hbox{E}\kern-.125emX}}
     \makeatletter
\newcommand{\linebreakand}{%
  \end{@IEEEauthorhalign}
  \hfill\mbox{}\par
  \mbox{}\hfill\begin{@IEEEauthorhalign}
}
\makeatother
    
\begin{document}

\title{Development of a Scaled Setup for Experimental
Study of the Effect of Lateral Dynamics on Energy
Consumption in Electric Vehicles: An Extension\\
 }


\author{Simran Kumari, Anand Ronald K., Siddhartha Mukhopadhyay and Ashish R. Hota
        
\thanks{Simran Kumari, Siddhartha Mukhopadhyay and Ashish R. Hota are with the Department of Electrical Engineering, IIT Kharagpur, West Bengal, India (e-mail: simranjnr@kgpian.iitkgp.ac.in; smukh@ee.iitkgp.ac.in; ahota@ee.iitkgp.ac.in). Anand Ronald K. is with the Advanced Technology Development Centre, IIT Kharagpur, West Bengal, India (e-mail: anandronald@outlook.com).}
}

\maketitle
\begin{abstract}
Most of the existing state-of-the-art approaches for energy consumption analysis do not account for the effect of lateral dynamics on energy consumption in electric vehicles (EVs) during vehicle maneuvers. This paper aims to validate this effect through an experimental study. We develop a scaled model using a radio-controlled (RC) car, modified to achieve dynamic similitude with on-road vehicles, to conduct scaled experiments. The experimental results confirm the impact of lateral dynamics on both energy demand and driving range in electric vehicles, aligning with our previous findings \cite{kumari2023energy}, and emphasize the need to incorporate these factors into energy consumption models. 

\end{abstract}


\section{Introduction}
Electric vehicles (EVs) have experienced a boom in the past few years due to being a sustainable alternative to vehicles with internal combustion engines. However, the issue of range anxiety among EV drivers has been persistent due to various reasons such as route, traffic, driver, and vehicle powertrain \cite{varga2019prediction}. Researchers in academia and industries have tried to address this issue through different approaches such as eco-driving \cite{shen2020minimum}, \cite{dib2014optimal}, eco-routing \cite{yi2018optimal}, \cite{chakraborty2021intelligent}, and range prediction \cite{varga2019prediction}, \cite{ondruska2014probabilistic}. 

In our previous work \cite{kumari2023energy}, we showed through simulation analysis that existing state-of-the-art approaches do not consider the effect of lateral dynamics on energy consumption during the maneuver and thus provide inaccurate results for eco-driving and range estimation. In this paper, we aim to validate the stated effect through experimental results.
 However, experimentation using full-size vehicles is extremely expensive and potentially dangerous. Several previous works in the automotive field have focused on developing scaled models for experimental verification and validation of intelligent control algorithms \cite{brennan2000illinois}, \cite{o2004scale}, \cite{hilgert2004development}, as well as electric powertrain emulation \cite{jeschke2013investigations}, \cite{neme2020}.
 The past work \cite{brennan2000illinois} designed a mechatronic testbed based on dynamic similitude analysis and the Buckingham $\pi$ theorem \cite{hanche2004buckingham} and validated the performance of a yaw-rate model compared to a reference model. Similarly, the work \cite{o2004scale} developed a scaled-model vehicle simulator to replicate the motion of a full-size vehicle. Furthermore, the paper \cite{milani2022importance} highlighted the significance of dimensional analysis for scaled vehicle-based experimentation for studying vehicle dynamics behavior. 

 In the domain of electric vehicles, the paper \cite{jeschke2013investigations} developed a Hardware-in-Loop (HIL) electric propulsion system setup to analyze the impact of traction systems on a vehicle's energy consumption.  Dimensional analysis is also employed during the design process. Furthermore, \cite{neme2020} used the Buckingham Pi theorem to design a motor-battery real-time simulation model that simulates the behavior of a light-duty powertrain setup.

Motivated by above past works, we develop a scaled model for our purpose. We use an RC car (RCC) and modify it to establish dynamic similitude with on-road vehicles. This is followed by designing and performing a scaled experiment that corresponds to the simulation analysis presented in the prior work \cite{kumari2023energy}. Additionally, we provide an energy consumption analysis based on the experimental results, including a discussion of the limitations of the present work. Finally, we present the conclusion listing future endeavors.

\section{Description of Scaled setup}
\begin{figure}[htbp]
    \centering
    \includegraphics[ width=0.75\linewidth,keepaspectratio]{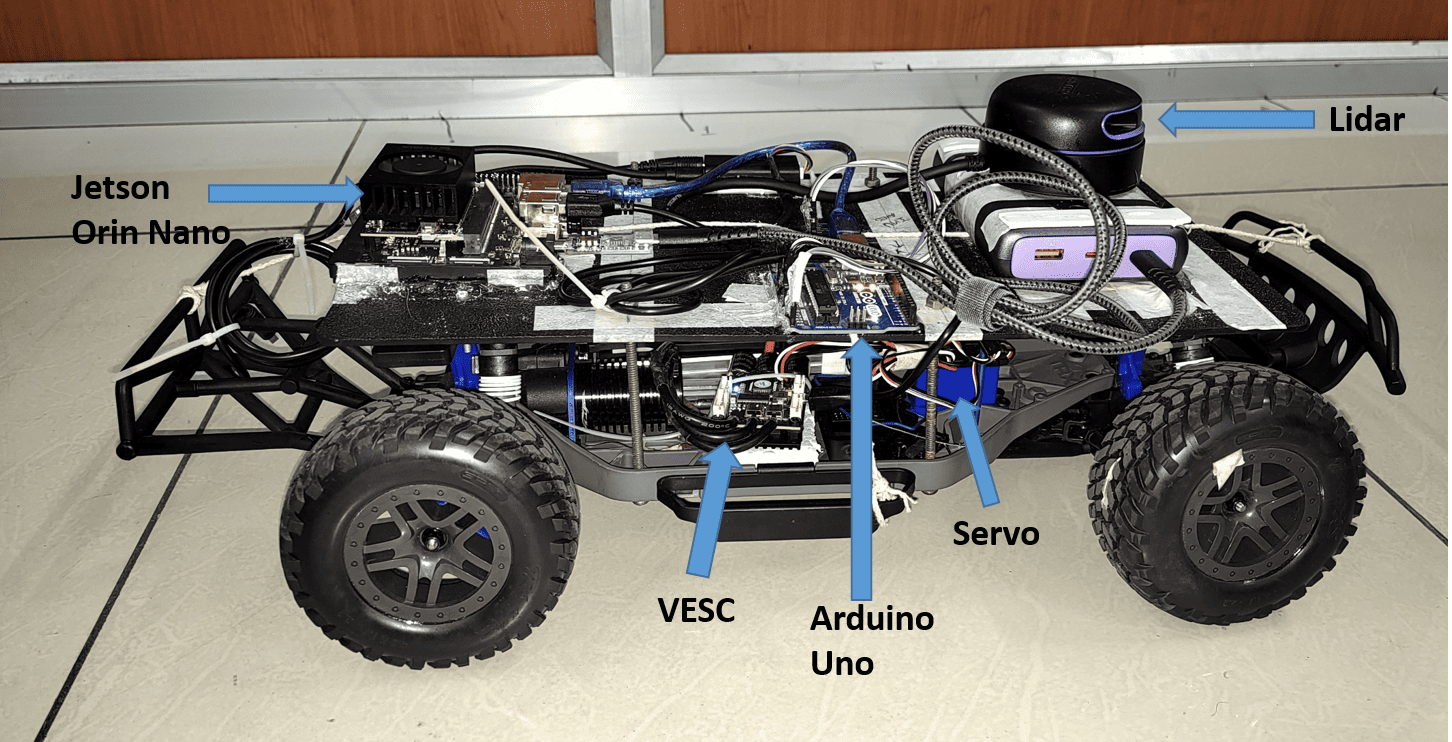}

   \includegraphics[ width=\linewidth,keepaspectratio]{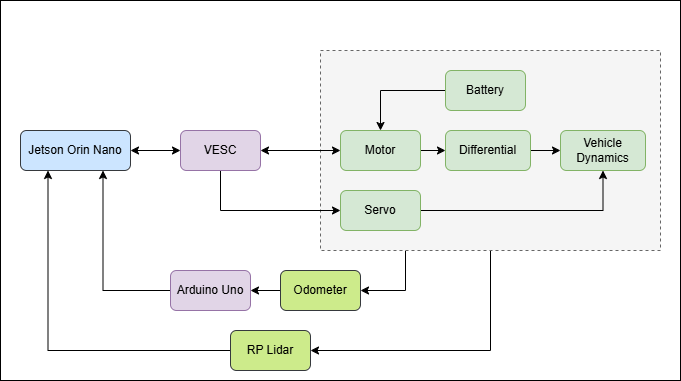}
    \caption{Details of RC car setup. }
    \label{fig:scematic}
\end{figure}

The scaled setup consists of a 1/10 four-wheel-driven Traxxas Slash electric RCC \cite{hu2019slasher} consisting of a 3S LiPo battery, a BLDC motor, a steering servo, a VESC motor controller \cite{choi2020development}, a PWM based servo controller, suspension, and a limited-slip differential assembly connected to wheels through a shaft. The setup is equipped with several sensors, including LIDAR and IMU, along with an Arduino Uno and a Nvidia Jetson Orin Nano for computation and processing. It is important to note that LIDAR measurements are not used for the experiments discussed in this work. Additionally, Table~\ref{tab:specs} provides specifications of the setup. The perception and control algorithms are built on ROS-based F1tenth driving stack \cite{o2020f1tenth}. 
  Fig. \ref{fig:scematic} shows the RC car and its corresponding connection diagram.  

 \begin{table}[htbp]
     \centering
      \caption{Specifications of the scaled setup}
     \begin{tabular}{|c|c|}
     \hline
         \multicolumn{2}{|c|}{\textbf{Battery}} \\ \hline
         Chemistry & LiPo \\ \hline
         Configuration & 3S \\ \hline
         Nominal Voltage & 11.1 V \\ \hline
         Nominal Capacity & 5.2 Ah \\ \hline
         Maximum Discharge Current & 40 C \\ 
         \hline
         \multicolumn{2}{|c|}{\textbf{Motor}} \\ \hline
          Rated Power& 800 W \\ \hline
          Rated Voltage & 12 V\\ \hline
          Rated Speed & 36300 rpm \\ \hline
          Rated KV & 3300 rpm/V \\ \hline
          Maximum speed & 50000 rpm \\ \hline
          \multicolumn{2}{|c|}{\textbf{Differential}} \\ \hline
          Gear ratio  & 11.82:1 \\ \hline
          \multicolumn{2}{|c|}{\textbf{Tires}} \\ \hline
          Type & Replica short course all-terrain\\ \hline
          Radius  & 49 mm \\ \hline
          \multicolumn{2}{|c|}{\textbf{ESC}} \\ \hline
          Input Voltage  & 8-60 V \\ \hline
          Input Continuous Current & 50 A \\ \hline
          Input Peak Current & 150 A \\ \hline
          BEC & 5V@1.5A \\ \hline
          ERPM Limit & 60000 \\ \hline
     \end{tabular}  
     \label{tab:specs}
 \end{table}
 \begin{table}[htbp]
     \centering
         \caption{Estimate parameters of RCC motor}
     \begin{tabular}{|c|c|}
     \hline
          Parameter& Value \\ \hline
          Motor resistance ($R$)&  19.5 m$\Omega$ \\ \hline
          Motor average inductance ($L_{avg}$) & 6.01 $\mu$H \\ \hline
          Motor inductance difference ($L_q-L_d$) & 1.31 $\mu$H \\ \hline
          Motor maximum flux linkage ($\lambda$)& 0.887 mWb \\ \hline
          Number of poles ($N_p$) & 4 \\ \hline
     \end{tabular} 
     \label{tab:mot_params}
 \end{table}
 
\textit{Motor Dynamics \& Control:} The Motor Control Unit (MCU), implemented using a VESC, regulates the operation of the BLDC motor based on a reference speed command provided by the supervisory controller. The MCU operates the motor in closed-loop speed control mode using a sensor-less Field-Oriented Control (FOC) scheme \cite{wu2017high}. This control method enables precise regulation of motor torque and speed by independently controlling the torque-producing and flux-producing current components in a rotating reference frame. The estimated motor parameters used in the control algorithm are listed in Table~\ref{tab:mot_params}.

In the FOC framework, the three-phase motor currents and voltages are transformed into a rotating direct–quadrature ($dq$) reference frame aligned with the rotor magnetic field using Clarke and Park transformations. In this reference frame, the $d$-axis component primarily controls the magnetic flux, while the $q$-axis component directly controls torque production. This decoupled control structure allows independent regulation of motor torque and flux, resulting in improved dynamic response, higher efficiency, and smooth operation compared to conventional control methods.

Assuming high efficiency of the inverter within the VESC, the electrical input power to the motor can be approximated as the battery power ($P_b$). The motor input power ($P_{in}$) in the $dq$ reference frame is given by
\begin{align}
P_b \approx P_{in} = \frac{3}{2}(V_q I_q + V_d I_d),
\end{align}

where $V_d$ and $V_q$ represent the direct-axis and quadrature-axis voltage components, respectively, and $I_d$ and $I_q$ represent the corresponding current components in the rotating reference frame. The factor $\frac{3}{2}$ arises from the power-invariant transformation from the three-phase stationary reference frame to the rotating $dq$ frame.

The electromagnetic torque generated by the motor ($\tau$) is expressed as
\begin{align}
\tau = \frac{N_p}{2} \frac{3}{2} \lambda I_q,
\end{align}

where $N_p$ denotes the number of motor poles, $\lambda$ represents the permanent magnet flux linkage, and $I_q$ is the quadrature-axis current component responsible for torque production. This relationship shows that motor torque is directly proportional to the $q$-axis current, which is regulated by the FOC controller to achieve the desired torque or speed response.

In sensor-less operation, the rotor position and speed are estimated internally by the VESC using back-electromotive force (back-EMF) and observer-based estimation techniques, eliminating the need for physical position sensors. The FOC controller uses this estimated rotor position to perform coordinate transformations and regulate current components accordingly. This control architecture enables efficient, smooth, and accurate motor operation, while also providing access to internal voltage, current, speed, and power signals required for energy flow analysis and experimental validation.




\section{Dynamic Similitude Analysis} \label{similitude_analysis}
 Scaling the behavior of the RCC to an actual vehicle on the road requires establishing a similarity between the two. This can be achieved using the concept of Buckingham $\pi$ Theorem from dimensional analysis \cite{hanche2004buckingham}. The theorem states that if there is a physically meaningful equation involving a certain number $n$ of physical variables, then the original equation can be rewritten in terms of a set of $p = n- k$ dimensionless parameters $\pi_1, \pi_2, \ldots, \pi_p$ constructed from the original variables, where $k$ is the number of basic physical dimensions involved. According to the theorem, two physical phenomena are similar if the values of the dimensionless parameters are the same for both.

\begin{figure}[]
\centerline{\includegraphics[width=0.45\textwidth,keepaspectratio]{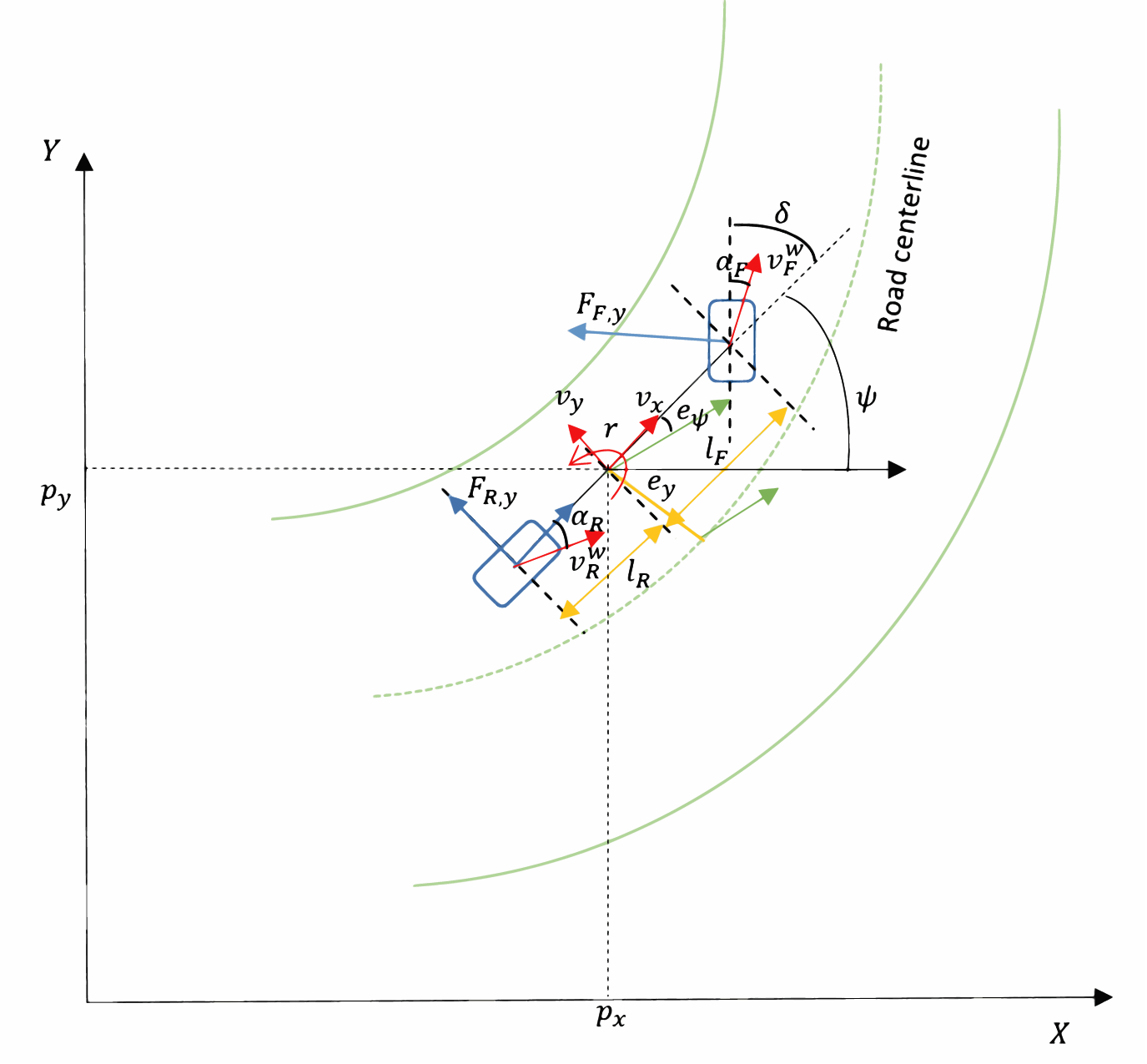}}
\caption{\footnotesize Schematic of different kinematic and dynamic variables associated with the bicycle model.}
\label{model}
\end{figure}

The behavioral characterization of both the RCC and on-road electric vehicles depends on vehicle and powertrain dynamics. However, we can neglect the transient behavior of the powertrain as its dynamics are much faster compared to the vehicle dynamics. The effect of lateral dynamics on energy consumption in electric vehicles, particularly when performing planar maneuvers, can be analyzed using the bicycle model \cite{kumari2024data} as given below:
\begin{align}
    \dot{v_x}&=a-\left ( \frac{F_{F,y}\sin \delta -m v_y r}{m}\right), \label{dyn_start} \\
    \dot{v_y}&= \frac{(F_{F,y}\cos \delta +F_{R,y}-m v_xr)}{I_z},\\
    \dot{r}&= \frac{(F_{F,y}l_F\cos \delta - F_{R,y} l_R)}{I_z},\\
    \tau&= \frac{r_wF_x}{\eta_i N_d}, \quad \omega= \frac{N_d}{r_w} v_x, \quad
    P_b=\eta_m \tau \omega,\label{dyn_end}\\
    \text{where, }& F_{F,y}=-2C_{F}\alpha_F, \quad F_{R,y}=-2C_{R}\alpha_R.
\end{align}
Here, the states $v_x$, $v_y$ are velocities along the longitudinal and lateral axis of the vehicle, and $r$ is yaw rate of the vehicle. These states evolve according to the above equations under longitudinal acceleration acting at the center of gravity (CoG) of the vehicle ($a$) and steering ($\delta$) as inputs. The corresponding schematic of the bicycle model and notation used for different parameters are provided in Fig \ref{model} and Table \ref{tab:notations} respectively. The dynamical equations in \eqref{dyn_start}-\eqref{dyn_end} results in a battery energy consumption ($E_b$) such that
\begin{align}
    E_b=\int P_b \,dt=\int \eta_m \tau_m \omega_m \,dt= \int \eta m a v_x\,dt. \label{stat_energy}
\end{align}
Here, we combine the motor efficiency ($\eta_m$) and differential efficiency ($\eta_i$) into a single powertrain efficiency parameter ($\eta= \frac{\eta_m}{\eta_i}$). 
\begin{table}[]
    \centering
    \caption{List of parameters for RC car and Rivian R1T}
\begin{tabular}{|p{4cm}|p{1.75cm}|p{1.75cm}|}
    \hline
     Parameters    & RCC& Rivian R1T \\ \hline \hline
      Mass, $m$ (kg) & 3.78 &  3152  \\\hline
      Yaw moment of inertia,$I_z$ (kgm$^2$) & 0.0382 & [4000, 6000]\\\hline
        Longitudinal distance of COG to front axle, 
        $l_F$ (mm) & 158.76 & 1691.5 \\\hline
      Longitudinal distance of COG to rear axle, $l_R$ (mm)& 165.23 & 1760.5 \\\hline
      Tire radius, $r_w$ (mm)& 49 & 419.1  \\\hline
      Differential gear ratio, $N_d$ & 11.82& 12 \\\hline
      Wheel base, $l$ (mm)& 324& 3452 \\\hline
      Front tire corner stiffness,         $C_F$ (N/rad)&  90& [40,000,50,000]\\\hline
       Rear tire corner stiffness,         $C_R$ (N/rad)&  90& [40,000,50,000] \\\hline 
        Rated motor power, $P_{m,\text{rated}}$ (W) & 800 & 400,000\\\hline 
        Rated motor torque, $\tau_{\text{rated}}$ (Nm)& 0.211 & 1231 \\\hline 
        Effective inertia of motor shaft, $J$ (kgm$^2$)& $2.2 \cdot 10^{-6}$ & $\approx$ 0.2 \\\hline
        Effective torsional damping of motor shaft, $B$ (Nm$\cdot$s/rad)& $1.17\cdot 10^{-5}$ & $\approx$ 0.1\\\hline\hline      
     Motor torque, speed, power  & \multicolumn{2}{|c|}{$\tau, \omega, P_m$} \\\hline
       Battery output power  & \multicolumn{2}{|c|}{$P_b$} \\\hline
        Lateral force at front and rear wheel  & \multicolumn{2}{|c|}{$F_{F,y}, F_{R,y}$} \\\hline
    \end{tabular}   
    \label{tab:notations}
\end{table}

There are seven independent variables in the model: $v_x$, $v_y$, $r$, $t$, $a$, $\delta$, $E_b$ as well as nine independent constants: $m$, $I_z$, $C_F$, $C_R$, $\eta$, $l$, $l_F$, $J$, $B$. Considering three basic physical dimensions, mass, length, and time, results in 13 dimensionless $\pi$ groups \cite{milani2022importance} corresponding to the physical behavior model \eqref{dyn_start}-\eqref{stat_energy} along with \eqref{drivetrain_dyn}, namely
\begin{align}
    \pi_1&=\delta, \quad \pi_{2}=\frac{ma}{C_F}, \quad \pi_3=\frac{l_F}{l}, \quad \pi_4={\frac{1}{C_Fl}}E_b, \nonumber\\ \pi_5&= \eta, \quad \pi_6= \frac{I_z}{ml^2},  \quad \pi_7= \frac{C_R}{C_F},\quad  \pi_8=\sqrt{\frac{C_F}{ml}}t \nonumber\\\pi_9&=\sqrt{\frac{m}{C_Fl}}v_x, 
     \quad  \pi_{10}=\sqrt{\frac{m}{C_Fl}}v_y, \quad \pi_{11}=\sqrt{\frac{ml}{C_F}}r,\nonumber\\
     \pi_{12}&={\frac{J}{ml^2}} , \quad \pi_{13}={\frac{B}{\sqrt{C_F m l^3}}}.
     \label{pi-groups}
\end{align}

The dynamic similarity between the RCC and an on-road electric vehicle requires that these $\pi$ groups be matched. Therefore, we modify the scaled setup to resolve the difference between the $\pi$ groups as described below.

Since the assembled RCC is based on a short-course truck model, the study focuses on validating the energy consumption behavior of typical commercial electric pick-up trucks. As a reference vehicle, we choose the luxury pick-up truck, Rivian R1T \cite{berk2020future}, and modify the RCC so that their $\pi$ groups are closely matched.

\begin{figure}
    \centering \includegraphics[width=0.75\linewidth, keepaspectratio]{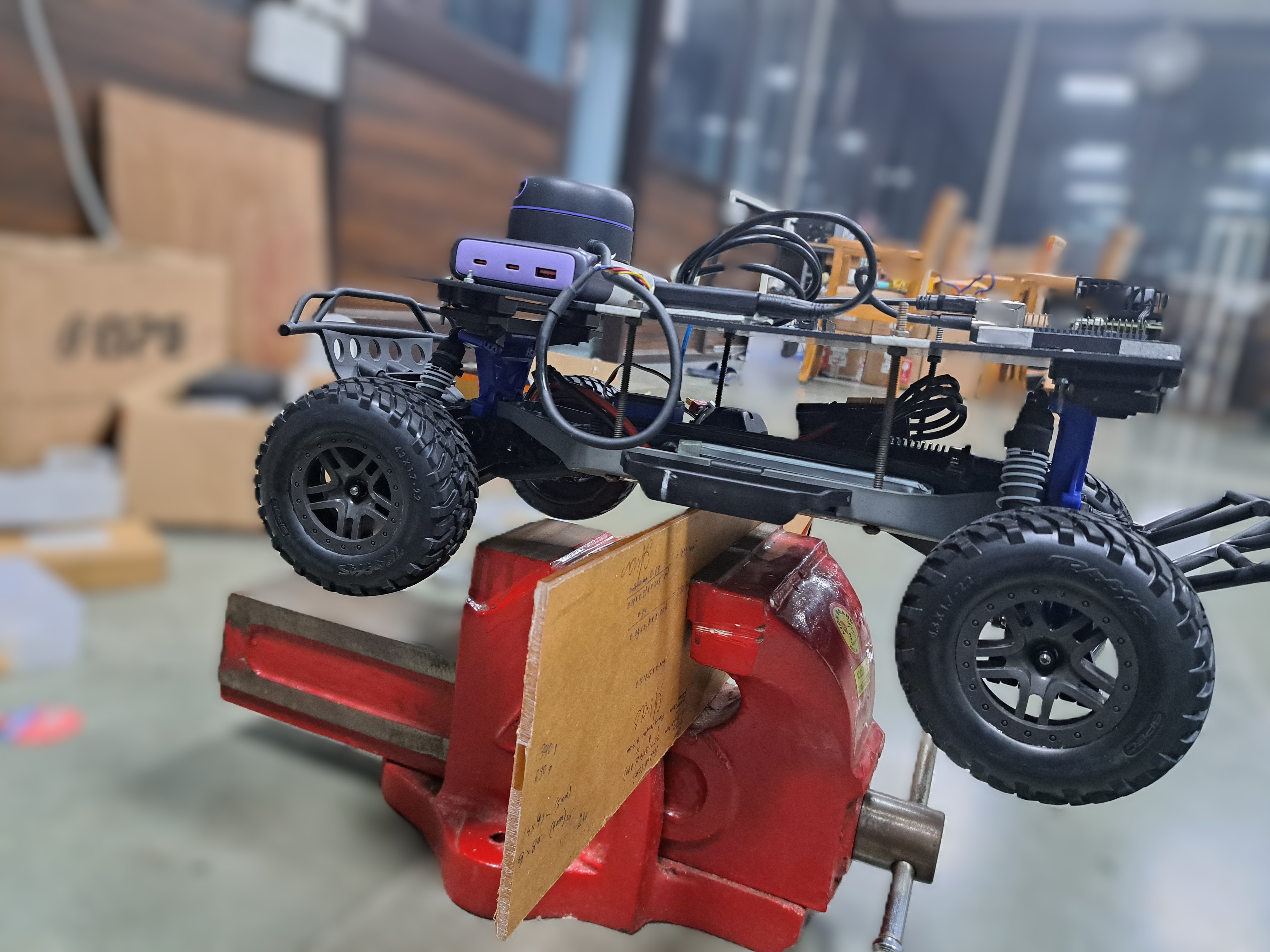}
    \caption{Mass balance setup for CoG calibration.}
    \label{fig:cog_pic}
\end{figure}

To calibrate the parameters $l_F$ and $l_R$, we adjust the CoG of the RCC setup. The mass distribution is performed by balancing the RCC over an acrylic sheet, which is secured using a vise, as shown in Fig. \ref{fig:cog_pic}. The balancing ensures that the resultant torque due to gravity forces is zero, allowing us to set the lengths $l_F$ and $l_R$ to 158.76 mm and 165.23 mm, respectively.

\begin{figure}[]
    \centering
    \includegraphics[height=6cm, keepaspectratio]{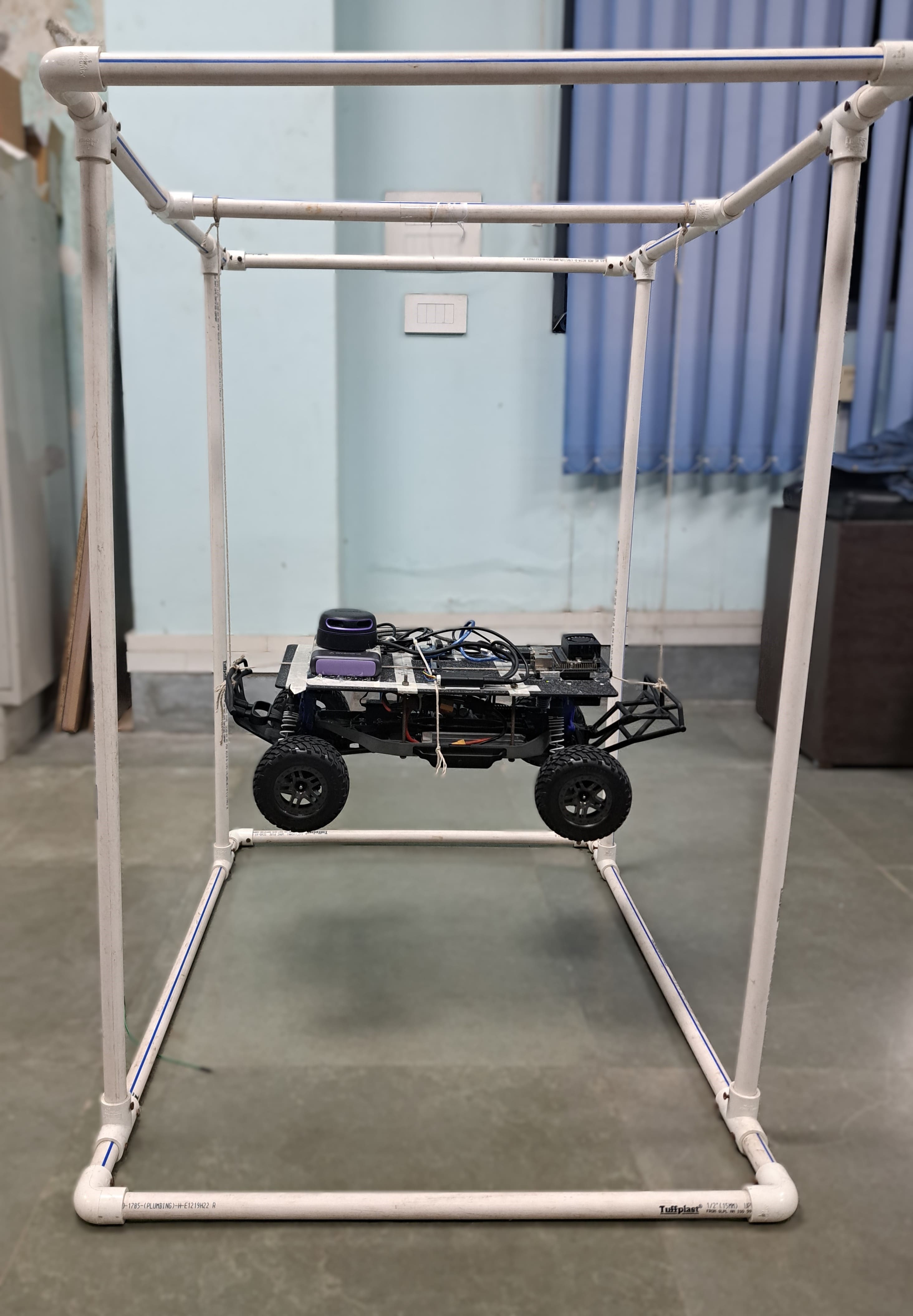}
    \caption{Bifilar suspension setup.}
    \label{fig:inertia}
\end{figure}

Next, we determine the moment of inertia ($I_z$) of the RCC setup using a bifilar suspension experiment \cite{jardin2009optimized}, as shown in Fig \ref{fig:inertia}. Based on the period of oscillations, we calculate inertia as per:
\begin{align}
    I_z=\frac{mgd^2T^2}{16\pi^2L}, \label{inertia}
\end{align}
where $d=53.5$ cm is the distance between parallel wires, $L= 60$ cm is the length of suspension wire, $g$ is the acceleration due to gravity, and $T= 0.589$s is period of oscillation. The value of inertia is determined to be 0.0382 kgm$^2$.

 Since the values for the moment of inertia and tire lateral stiffness are not available for Rivian R1T, we use typical values observed for pick-up trucks. 
 The normalized yaw moment of inertia, defined as $I_z/(m\cdot(l/2)^2)$, typically ranges from $0.7$ to $0.9$ for pickup trucks \cite{heydinger1999measured}. The yaw moment of inertia is slightly less than this value and lies within $[4000, 6000]$ kgm$^2$. Further, the cornering stiffness lies within $[40,000,50,000]$ N$/$rad \cite{han2018estimation} .
We assume an approximate value of tire stiffness for the RCC \cite{brennan2000illinois}. The specifications and corresponding $\pi$-group values for the RCC and the Rivian R1T are provided in Table \ref{tab:notations} and \ref{tab:pi_list}, respectively.  
 We observe that the values of constant $\pi$ groups, with the exception of $\pi
 _5$, $\pi_6$ and $\pi_{13}$, are nearly identical for both vehicles. It is important to note that the remaining $\pi$ groups are functions of time and scales according to \eqref{pi-groups}. Thus, according to (\ref{pi-groups}) and Table \ref{tab:pi_list}, the variables corresponding to the RCC on an average scale are as follows:
\begin{equation}
        (v_x)_{RC}= 0.416 (v_x)_{EV},
    \quad (t)_{RC}= 0.248(t)_{EV}.  \label{vxt}
\end{equation}
This analysis suggests that the dynamic behavior of the RC car can reasonably predict the dynamic behavior of an on-road vehicle similar to Rivian R1T, provided the vehicle exhibits RCC-compatible motor efficiency, yaw moment of inertia, and relatively high drivetrain torsional damping  (around 0.25 Nm$\cdot$s/rad). A detailed discussion of the differences between the RCC and the Rivian R1T is provided in the Appendix~\ref{appendix1}.
\begin{table}[htbp]
    \centering
    \caption{List of constant $\pi$ group values}
    \begin{tabular}{|p{2cm}|p{2cm}|p{2cm}|}
    \hline
    $\pi$ groups & RCC & Rivian R1T\\ \hline
        $\pi_3$  & 0.51& 0.51\\
        $\pi_5$  &varying (Fig. \ref{rcc_mot_map}) & varying (Fig. \ref{rivian_mot_map})\\
        $\pi_6$  & 0.0983& [0.106, 0.159] \\
        $\pi_7$  & 1 & 1\\ 
        $\pi_{12}$  & $5.54\cdot10^{-6}$ & $\approx 5.35\cdot10^{-6}$\\
        $\pi_{13}$  & $3.44\cdot10^{-6}$ & $\approx 1.37\cdot10^{-6}$\\ \hline
    \end{tabular} 
    \label{tab:pi_list}
\end{table}

\section{Effect of maneuver on Battery energy consumption}
\begin{figure}[htbp]
    \centering
    \includegraphics[width=\linewidth, height=5cm,keepaspectratio]{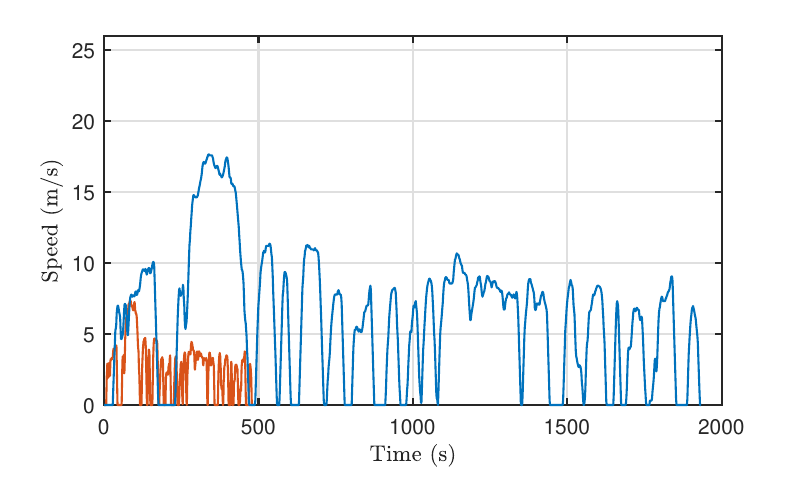}
    \caption{Actual (blue) and scaled (red) drive cycle used for experimental study.}
    \label{fig:udds}
\end{figure}

Supported by similitude analysis, we conducted experiments with the RCC to study the behavior of the commercial electric pick-up trucks similar to the Rivian R1T after including stated modifications. Considering the speed and acceleration limitation, the RCC is made to track a scaled drive cycle shown in Fig. \ref{fig:udds}. The blue curve in Fig. \ref{fig:udds} represents the equivalent on-road drive cycle profile. Scaling the drive cycle according to \eqref{vxt} results in a scaling of the traveled distance approximately by a factor of 9.38, which is reasonable for the setup under study.

\begin{figure}[htbp]
    \centering
    \includegraphics[width=0.75\linewidth, keepaspectratio]{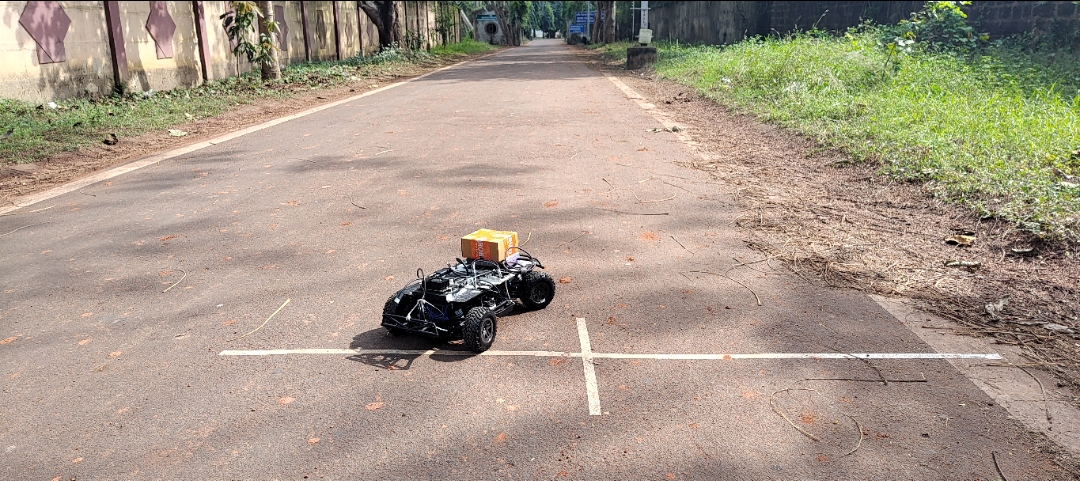}
    \caption{Test track for experimentation.}
    \label{fig:track}
\end{figure}

 With this scaling a lane change approximately every 50 m is reasonable \cite{kumari2024data}. A straight road being available was approximately 150 m in length as shown in Fig. \ref{fig:track}. Therefore, the effect of lateral dynamics on energy consumption is demonstrated by performing lane changes every 20 m on this road track.

\begin{figure}[htbp]
    \centering
    \includegraphics[width=\linewidth, keepaspectratio]{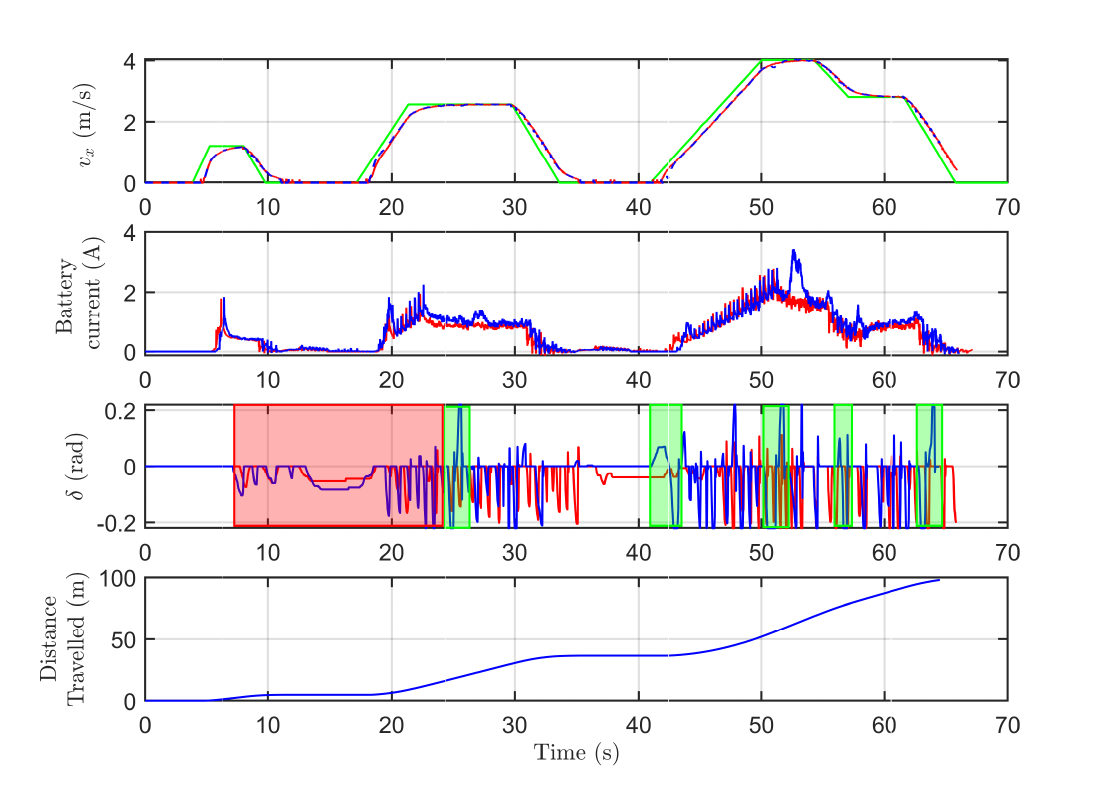}
    \centering
    \includegraphics[width=\linewidth, keepaspectratio]{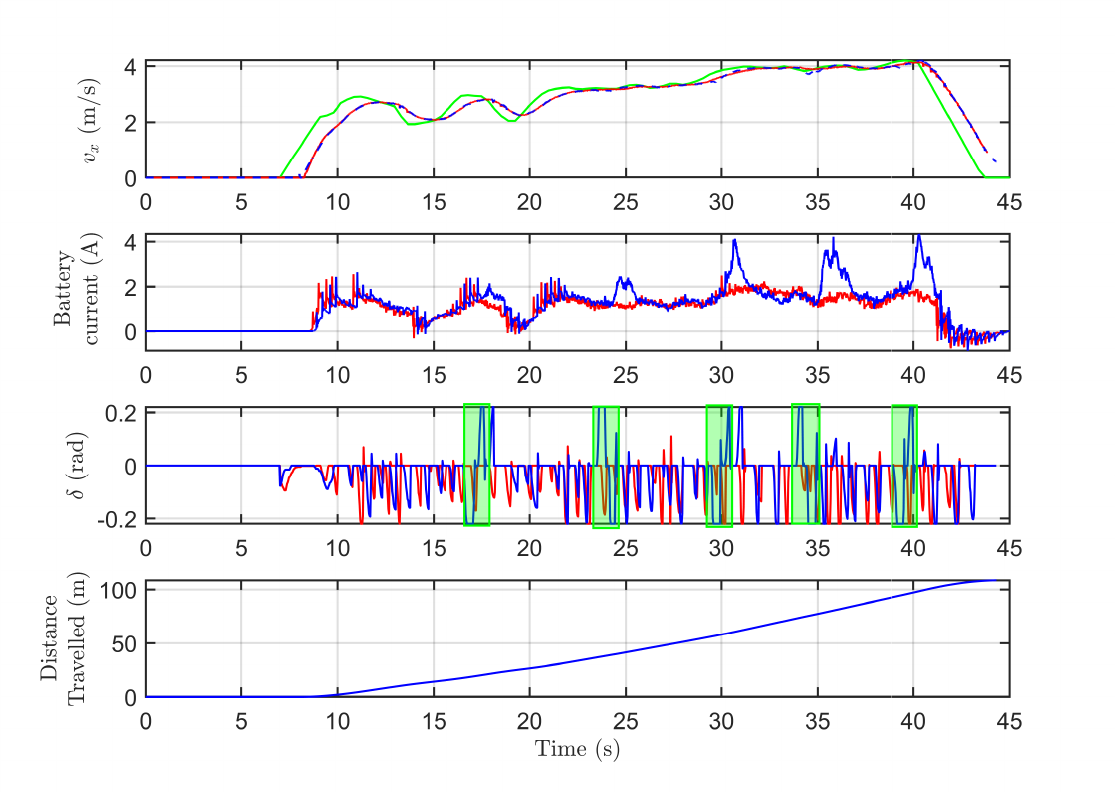}
    \caption{Parameter variation profile for scaled drive cycles. Drive cycles in (a) and (b) are referred to as Scaled 1 and Scaled 2, respectively. Red: without lane change, Blue: with lane change. The green box in the third subplot highlights the lane change instant.}
    \label{fig:exp_plot}
\end{figure}

As shown in Fig. \ref{fig:exp_plot}, a lane change increases current demand while tracking the same speed profile as highlighted by the steering profile in the green box. Frequent steering inputs are required to counteract the vehicle's drifting (highlighted by a red box), in addition to the lane changes. It is important to note that the RCC is not equipped with brake pads and does not leverage regenerative braking at the high sampling rate of 100 Hz, which is commonly employed in on-road EVs. The experimental data corresponding to Fig. \ref{fig:exp_plot} is summarized in Table \ref{tab:data_exp}, showing an increase in energy consumption for drive cycles referred to as Scaled 1 and Scaled 2, respectively, compared to straight driving.

 To ensure a fair comparison of battery energy consumption between the RCC and a full-scale vehicle, we simulated an EV model, similar to the Rivian R1T, within Simulink. This model is modified to incorporate RCC-compatible motor efficiency, yaw moment of inertia, and drivetrain torsional damping. 
 The resulting energy consumption data is listed in Table~\ref{tab:data_sim}. The simulated and actual (experimental) energy efficiency values for the RCC are presented in Table~\ref{tab:energy_eff}. Based on the scaling relationship in Equation~\eqref{pi-groups}, a battery energy consumption of 1 kWh/km in a full-size EV with $C_F/C_R=40,700$ N/rad corresponds to approximately 0.0019 Wh/m for the RCC. The observed difference between the expected and actual scaled energy efficiency ranges from 8-25$\%$, which is considered acceptable. 

     \begin{table}[hbtp]
     \centering
      \caption{Measurement Data for RCC: Energy (Wh) and Distance (m)}
        \centering

         \begin{tabular}{|p{1.75cm}|p{1cm}|p{1cm}|p{1cm}|p{1cm}|}
    \hline
         \multirow{2}{*}{Drive cycle}& \multicolumn{2}{|c|}{Without lane changes} & \multicolumn{2}{|c|}{With lane changes} \\ \cline{2-5}
         & Energy  & Distance  & Energy  & Distance\\\hline
         Scaled 1 & 0.1265 & 99.16 & 0.1469 & 98.02 \\ \hline
         Scaled 2 & 0.1427 & 108.29 & 0.1709 & 108.65 \\ \hline
    \end{tabular}
        \label{tab:data_exp}
    \end{table}
 \begin{table}[hbtp]
 \centering
   \caption{Simulation Data for EV: Energy (kWh) and Distance (km) }
    \begin{tabular}{|p{1.25cm}|p{0.75cm}|p{0.8cm}|p{0.8cm}|p{0.75cm}|p{0.8cm}|p{0.8cm}|}
    \hline
         \multirow{2}{*}{Drive cycle}& \multicolumn{3}{|c|}{Without lane changes} & \multicolumn{3}{|c|}{With lane changes} \\ \cline{2-7}
         & Energy  & Distance  & kWh/km& Energy  & Distance  & kWh/km\\\hline
         1 & 0.7486 &0.992 & 0.7546 & 0.8375 & 0.991 & 0.8451\\\hline
         2 & 0.9144 &1.068& 0.8562 & 1.0842 & 1.069 & 1.0142\\\hline
    \end{tabular}
    \label{tab:data_sim}
    \end{table}
    
\begin{table}[hbtp]
\centering
   \caption{Energy efficiency of RCC in Wh/m. ({A battery energy consumption of 1kWh/km for EV scales to 0.0019Wh/m in RCC.})}
     \begin{tabular}{|p{1.75cm}|p{1.25cm}|p{1.25cm}|p{1.25cm}|p{1.25cm}|}
\hline
         \multirow{2}{*}{Drive cycle}& \multicolumn{2}{|c|}{Without lane changes}  & \multicolumn{2}{|c|}{With lane changes}\\ \cline{2-5}
      & Actual  & Simulated & Actual  & Simulated\\\hline
        Scaled 1 & 0.00127&0.00143& 0.00149 & 0.00161 \\\hline
        Scaled 2 & 0.00132 &0.00163& 0.00157 & 0.00193\\\hline
 \end{tabular}
 \label{tab:energy_eff}
    \end{table}

The Rivian R1T has a rated energy efficiency of 48 kWh/100 mi for the combined city/highway driving cycle, which scales to 0.00057 Wh/m for the RCC. This value deviates significantly from the experimental RCC energy efficiency. However, when a lower drivetrain torsional damping (around 0.1 Nm$\cdot$s/rad) and the motor efficiency shown in Fig. \ref {rivian_mot_map} are considered, the discrepancy is reduced to within 22\%, yielding a scaled energy consumption of 0.00074 Wh/m. These adjustments suggest that the scaled setup can predict the energy behavior of a vehicle similar to the Rivian R1T with reasonable accuracy. Furthermore, Table \ref{tab:lc_energy} indicates that energy efficiency increases by 3.46\% and 3.78\% for maneuvers in the presence of lane changes. The simulated trends exhibit comparable values and align with the finding reported in \cite{kumari2023energy}. 

\begin{table}[htbp]
    \centering
        \caption{Energy efficiency difference  for maneuvers with and without lane changes}
    \begin{tabular}{|p{1.75cm}|p{1.25cm}|p{1.25cm}|} \hline
         \multirow{2}{*}{Drive cycle}& \multicolumn{2}{|c|}{\% Increase in energy efficiency}  \\ \cline{2-3}
         & Actual & Simulated  \\ \hline
        Scaled 1 & 3.46 & 3.14 \\ \hline
        Scaled 2 & 3.78 & 3.68 \\ \hline
    \end{tabular}
    \label{tab:lc_energy}
\end{table}


In summary, the developed scaled setup, supported by similitude analysis, is capable of predicting the driving behavior of a vehicle similar to the Rivian R1T with reasonable accuracy. Moreover, it effectively illustrates the impact of lateral dynamics on energy consumption.

\section{Conclusion}
In this work, we develop a scaled model of an EV by modifying an RC car and establishing dynamic similitude with on-road vehicles. This is followed by the design and execution of a scaled experiment, accompanied by a thorough analysis, to study the effect of lateral dynamics on energy demand in electric vehicles. The experimental findings highlight that state-of-the-art range prediction and eco-driving approaches tend to overestimate energy consumption, motivating the need to include lateral dynamics in these models.


\section*{Acknowledgment}

The authors would like to thank Mr. Avik Mazumdar for his assistance in conducting the experiments, and Mr. Kamal Sandeep Karreddula for providing valuable technical inputs.

\appendices

\section{Description of Factors Affecting Similitude Analysis} \label{appendix1}
In our earlier work \cite{kumari2025development}, we presented a method of developing a scaled setup using an RCC that is similar to on-road vehicles. The experimental results demonstrated increased energy consumption during lateral maneuvers, such as lane changes. However, the validation of dynamic similarity exhibited inaccuracies due to several factors that differentiate the RCC from a full-scale EV.   

The sources of the stated inaccuracy are discussed below:
\begin{enumerate}
\item \textbf{Torsional Damping of the drivetrain: } In the original analysis, powertrain dynamics were not incorporated into the similitude framework for evaluating energy consumption. However, energy transfer from the propulsion source (i.e., the motor) to the wheels is influenced by the rotational inertia and torsional damping of the drivetrain.
The motor shaft dynamics can be represented by the following equation:
\begin{equation}
    J\dot{\omega}+B\omega+T_L=\tau, \label{drivetrain_dyn}
\end{equation}
where $J$ and $B$ denote the effective inertia and torsional damping at the motor shaft, accounting for reflected inertia and damping from drivetrain components and wheels. \footnote{For a 4WD EV, the effective inertia can be expressed as
\begin{align*}
    J=J_m+J_{drive shaft}+J_{differential}+4\frac{J_{axle}}{N_d^2}+4\frac{J_{wheel}}{N_d^2}.
\end{align*}. A similar relation can be used to estimate the effective torsional damping.}
The load torque $T_L$ represents the load torque due to tire-road interaction, as reflected on the motor shaft. Hence, part of the propulsive power is utilized to overcome the effects of inertia and damping in the drivetrain. 
\begin{table}[htbp]
    \centering
    \caption{RCC Powertrain behavior}
    \begin{tabular}{|p{4cm} p{2cm}|}
    \hline
    Observed quantities & Value \\ \hline
        Steady-state motor speed ($\omega_s$)&  527 rad/s\\
         Steady-state motor torque ($\tau_s$)&  0.0067 Nm\\
         Time constant ($T_s$) & 0.196 s\\ \hline
    \end{tabular}
    \label{tab:motor_behavior}
\end{table}
These parameters were obtained by applying a step speed reference to the motor while the RCC wheels spun freely. From Table~\ref{tab:motor_behavior}, the estimated parameters are:
\begin{align*}
    B&=\frac{\tau_s}{\omega_s}=1.17\cdot10^{-5} \quad \text{Nm}\cdot \text{s/rad}, \nonumber \\ 
    J&=B\cdot T_s=2.2\cdot10^{-6} \quad \text{kgm}^2.
\end{align*}

\item \textbf{Motor efficiency mismatch: } In typical on-road EVs, motor efficiency ranges between 85 and 97\%, as illustrated in Fig.~\ref{rivian_mot_map}. In contrast, the RCC motor shows significantly lower efficiency based on measured data, as shown in Fig.~\ref{rcc_mot_map}. This discrepancy invalidates the assumption of constant efficiency incorporated in the dimensionless parameter $\pi_5$. Further, the RCC must operate within the scaled torque-speed envelope (shown in cyan) in Fig~\ref{rcc_mot_map} to maintain similarity with an on-road vehicle similar to Rivian R1T.
\begin{figure}[t]
    \centering

    \begin{subfigure}{0.5\textwidth}
        \centering
        \includegraphics[height=2in]{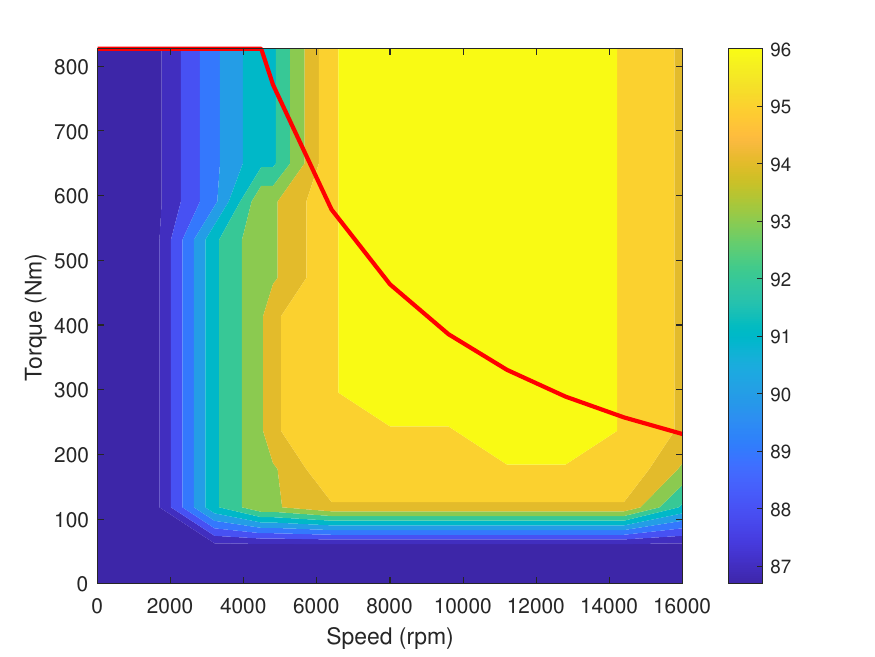}
        \caption{Motor efficiency map for Rivian R1T 400 kW Dual-motor}
        \label{rivian_mot_map}
    \end{subfigure}

    \bigskip

    \begin{subfigure}{0.5\textwidth}
        \centering
        \includegraphics[height=2in]{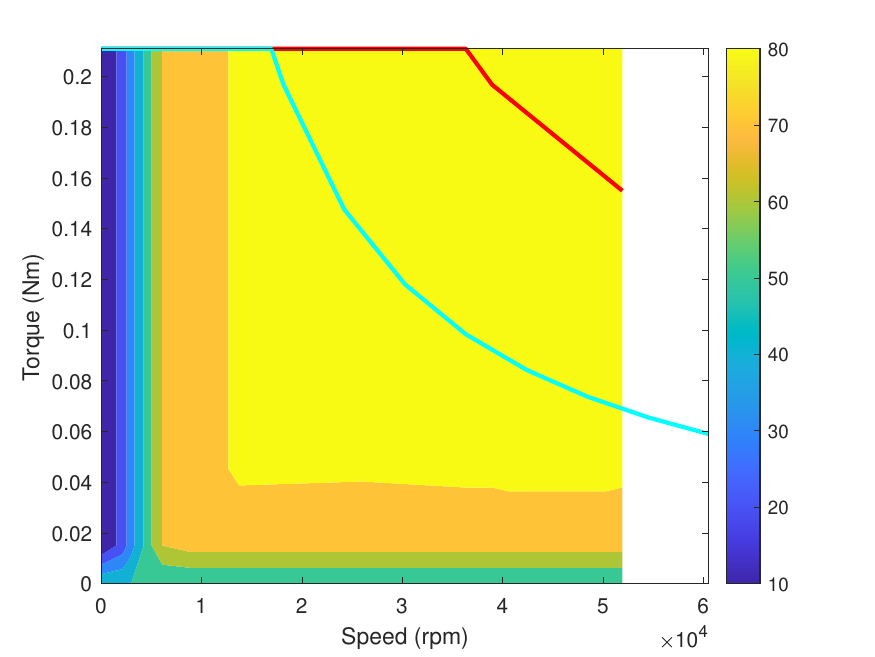}
        \caption{Estimated motor efficiency map for RC car 800 W motor}
        \label{rcc_mot_map}
    \end{subfigure}

    \caption{Motor efficiency map with torque-speed characteristics. Red: actual torque-speed curve, Cyan: Expected torque-speed curve according to Rivian R1T.}
    \label{fig:mot_map}
\end{figure}




    \item \textbf{Discrepancy in acceleration scaling: } The dimensionless parameter $\pi_2$, introduced in \eqref{pi-groups}, represents the scaling of force for dynamically similar phenomena. 
Although the RCC operates with a high control loop frequency (1000 Hz) and a high sampling frequency for the reference speed (approximately 100 Hz), these discrete-time control settings fail to replicate the continuous-time behavior of an EV driver model. Therefore, achieving proper force scaling requires the generation of a discrete-time reference speed profile in a simulation environment to enable accurate validation.

\textbf{In the context of regeneration:} For on-road EVs, the closed-loop control architecture typically involves the driver issuing acceleration or deceleration commands, which are interpreted by the Vehicle Control Unit (VCU). The VCU processes these inputs and generates a reference motor torque command, which is then executed by the MCU. Thus, inside an EV, a motor operates in torque control mode, meaning even slight deceleration inputs can lead to regenerative braking or mechanical braking. The actual outcome depends on several factors, including vehicle speed, battery state of charge (SOC), battery C-rate limits, and available motor torque.

In contrast, RCC employs a different control strategy. The MCU receives a reference speed command and operates the motor in speed control mode. Here, small deceleration is achieved by reducing the speed reference, thereby lowering the output of the internal speed control loop without invoking negative torque.

To effectively utilize regenerative braking in the RCC, the motor must be driven in a high-torque regime, which would require reducing the reference speed sampling rate. However, lowering the sampling rate causes oscillations and jerk in the RCC’s dynamics. The similar behavior would be seen in the corresponding EV model when tracking the scaled drive cycle, which is inconsistent with typical city or highway driving patterns. Therefore, a trade-off exists between regeneration capability and smooth acceleration. For this study, a higher sampling frequency of 100 Hz was chosen to preserve trajectory smoothness.

        \item \textbf{Speed control-loop dead zone at low speed: }It was observed that the RCC motor exhibits a dead zone at low speeds, requiring a minimum speed threshold of approximately 400 rpm to initiate motion. This leads to a torque and current spike when starting from rest, as illustrated  in Fig.~\ref{fig:exp_plot}. While this has a minor impact on total energy consumption, it can be incorporated into the simulation model by introducing a speed feedback dead zone at low speeds.
\end{enumerate}

Accounting for these factors improves the fidelity of the simulated model, aligning its behavior more closely with that of the RCC.

\bibliographystyle{ieeetr}
\bibliography{itec_extension} 

\end{document}